\documentstyle[12pt]{article}
\begin{document}
\newcommand{\vsone}{\vspace{1cm}}
\newcommand{\be}{\begin{equation}}
\newcommand{\ee}{\end{equation}}
\newcommand{\bea}{\begin{eqnarray}}
\newcommand{\eea}{\end{eqnarray}}
\newcommand{\pr}{\paragraph{}}
\newcommand{\nk}{\noindent}

\newcommand{\nd}[1]{/\hspace{-0.6em} #1}
\begin{titlepage}
\begin{flushright}
CERN-TH.6534/92\\
ACT-12/92 \\
CTP-TAMU-47/92 \\
\end{flushright}

\begin{centering}
\vspace{.1in}
{\Large {\bf World-Sheet Duality, Space-Time
Foam, and the Quantum Fate of a Stringy Black Hole }} \\

\vspace{.2in}
{\bf John Ellis}, {\bf N.E. Mavromatos} and {\bf D.V.
Nanopoulos}$^{\dagger}$   \\
\vspace{.05in}
Theory Division, CERN, CH-1211, Geneva 23, Switzerland  \\

\vspace{.05in}
\vspace{.05in}
\vspace{.1in}
{\bf Abstract} \\
\vspace{.03in}
\end{centering}
{\small
\paragraph{}
We interpret Minkowski black holes as world-sheet
{\it spikes } which are related by world-sheet
{\it duality}
to {\it vortices } that
correspond to Euclidean black
holes. These world-sheet defects
induce defects in the gauge fields
of the corresponding coset Wess-Zumino
descriptions of spherically-symmetric
black holes. The
low-temperature
target space-time foam is a Minkowski black hole
(spike) plasma
with confined Euclidean black holes (vortices).
The
high-temperature phase is a {\it dense} vortex plasma
described by a topological gauge field theory
on the world-sheet, which possesses enhanced symmetry
as in the target space-time singularity at the core of a black
hole. Quantum decay via higher-genus effects
induces a back-reaction which
causes a Minkowski
black hole to lose mass until it is indistinguishable
from intrinsic fluctuations in the space-time foam.}

\par
\vspace{0.4in}
\vspace{0.1in}
\begin{flushleft}
CERN-TH.6534/92 \\
ACT-12/92 \\
CTP-TAMU-47/92 \\
June 1992 \\

\vspace{0.4in}
$^{\dagger}$ {\it Permanent address} :
Center for Theoretical Physics, Dept. of Physics,
Texas A \& M University, College Station, TX 77843-4242, USA,
and  \\
Astroparticle Physics Group,
Houston Advanced Research Center (HARC),
The Woodlands, TX 77381, USA.\\

\end{flushleft}

\end{titlepage}
\newpage
\section{Introduction and Summary}
\pr
    The reconciliation of gravity with quantum mechanics raises
problems at several different levels. One is the meaningful
calculation of quantum gravitational corrections to scattering
processes in a flat space-time background. Another is the treatment of
quantum effects in a topologically non-trivial background, such as a
black hole, with an event horizon and a singularity. Then there is
the problem of treating microscopic quantum gravitational
fluctuations in the space-time vacuum, i.e., space-time foam. Next
there is the phase structure of space-time, and the possibility that
a new, more symmetric phase may appear at high temperatures and/or
close to singularities. And so on ....
\pr
    String theory is sufficiently ambitious to claim to solve these
and all other problems of quantum gravity. So far, progress has been
more modest. However, it has been shown that string theory gives
finite results in each order of perturbation theory \cite{mand}, so
that the quantum corrections to scattering in flat space-time are
indeed tamed. The discovery of a black hole solution of string theory
in a two-dimensional space-time \cite{witt}, which can equally well be
regarded as a spherically-symmetric solution in a four-dimensional
space-time \cite{emn4}, has opened the way to the study of quantum
effects in a topologically non-trivial space-time background. We have
shown that quantum coherence is maintained in a stringy black hole
background \cite{emn1}, thanks to an infinite set of
gauged \cite{emn1,ind}
W-symmetries \cite{whoW} and associated conserved charges
\cite{emn1}, which
we have called W-hair. We have also shown \cite{emntop} how a higher
double W-symmetry is realized at the singularity at the core of the
stringy black hole, and is broken down to the remaining exact
W-symmetry by an infinite generalization of the
Berezinsky-Kosterlitz-Thouless
\cite{KT} vortex condensation mechanism. The symmetric phase at
the singular core of the black hole is described by a topological
gauge field theory on the world-sheet \cite{witt,eguchi}, whilst the
symmetry-breaking corresponding to the generation of a non-singular
space-time metric is akin \cite{emntop} to the condensation of matter
fields such as quarks in QCD \cite{chiral}.
\pr
    In this paper we build on the above results to discuss the nature
of space-time foam and the phase structure of space-time. We show
that whereas world-sheet vortices (topological solitons)
correspond to Euclidean thermal black
holes, the dual world-sheet ``spikes'', which are
non-topological solitons
correspond to Minkowski
black holes. We discuss in both cases
the corresponding ``monopole'' configurations
of the associated gauge fields in their
respective Wess-Zumino coset model representations.
We then show that
space-time has a low-temperature phase in which the
vortices are
confined
and there is a plasma of
spikes on the
world-sheet, i.e., Minkowski black holes in target space. Any
macroscopic black hole decays via the quantum effects associated with
higher genera \cite{emn3}, its mass decreasing via back-reaction
until it is
indistinguishable from a quantum-mechanical fluctuation in the
low-temperature vacuum. There is a high-temperature phase
characterized by spike
confinement and a high-density plasma
of vortices, in which the higher stringy symmetry is restored as at the
core of a black hole. These results are derived explicitly in a
two-dimensional target space-time, but we give naive entropy arguments
suggesting
a similar phase structure for four-dimensional target
space-time.

\section{Dual Defects on the World-Sheet }
\pr
    In this section we introduce the dual pair of types of defect
that can appear on the world-sheet, namely vortices and
``spikes'' associated with ``monopoles'' in different
auxiliary gauge fields, and
display their respective interpretations as Euclidean and Minkowski
black holes in target space-time.
\pr
{\it Vortices}: These are solutions $X_v$ of Green function equations of
the following type \cite{abr,thom}:
\be
\partial _z {\overline \partial _z }
X_v =\frac{i\pi q_v}{2} [\delta(z-z_1)-\delta(z-z_2)]
\label{vorpair}
\ee

\nk which corresponds to a vortex centred at $z_1$ and an antivortex
centred at $z_2$, and we have made a stereographic projection of
the lowest-genus world-sheet sphere onto the complex $z$-plane.
There is always zero net vorticity on the compact world-sheet of
the closed string, so such defects always appear in vortex/antivortex
pairs. If the vortex is located at the origin (South Pole) and the
antivortex at infinity (North Pole), the
corresponding $\sigma$-model
coordinate solution of (\ref{vorpair}) is
\be
    X_v=q_v Imlnz
\label{vor}
\ee

\nk It is clear from this representation that in order for $X_v$ to be
single-valued, it must have period $2\pi$, and hence
the vortex
charge $q_v$ must be an integer.
\pr
    Since they require periodicity of the world-sheet scalar
field, vortices are topological solitons that
appear only in models with compactified
target space-time dimensions. The application that interests us is to
Euclidean black holes in equilibrium with a heat-bath, that are
described by a periodic imaginary time coordinate. In this
application, the charge $q_v$ of the vortex is proportional to the
square root of the black hole mass. The
vortex quantization condition
implies that the Euclidean black hole cannot lose any mass, as should
be expected from its equilibrium with a heat bath (reservoir).
\pr
To see formally the space-time
interpretation of a world-sheet vortex
as a Eucldidean
black hole, we rewrite the vortex configuration $\theta$
as
\be
      e^{2i\theta} =\frac{z}{{\bar z}}
\label{vort}
\ee

\nk and complexify the phase
by introducing a real part $r$, which is
defined by
the following
embedding of the world-sheet
in a two-dimensional target space spanned by
$r$ and $\theta$ \cite{chinese},
\be
  z=(e^r-e^{-r})e^{i\theta}
\label{emb}
\ee

\nk The induced target-space metric is inferred from
the world-sheet arc-length, which is
$dl=\frac{dz}{1+z{\bar z}}$  after the stereographic
projection,
and the infinitesimal
Euclidean displacement $d{\bar z}$ induced by the $r$-coordinate.
The result is a Euclidean black hole with a coordinate
singularity at $r=0,\theta=0$, which corresponds to the position of
the world-sheet vortex, and
\be
     ds^2_{target} \equiv \frac{dzd{\bar z}}{1+z{\bar z}}=
dr^2 + tanh^2 r d\theta^2
\label{ebh}
\ee

\nk is the target space-time line element.
\pr
Because of the
representation of the Euclidean black hole
as a coset $\frac{SL(2,R)}{U(1)}$ Wess-Zumino ($WZ$)
model \cite{witt},
the world-sheet vortices induce gauge
defects of the underlying world-sheet gauge
theory. The compactness
of the gauged $U(1)$ subgroup
and the non-triviality of its
first homotopy class $\Pi _1 (U(1))= {\bf Z}$
suggest
the identification of the vortex configurations with topologically
non-trivial gauge field configurations of the gauged $WZ$ model.
Indeed, eliminating the non-propagating gauge field $A$ of the
gauged $WZ$ model using its equations of motion
yields
\be
A_{z} =-\frac{u\partial _z (a-b)
- (a-b)\partial _z u  }{ (a+b)^2 }
\label{sing}
\ee

\nk and
a similar
expression for $-A_{\bar z}$ (with $z$ replaced everywhere
by ${\bar z}$).
Here, $ u = sinhr sin\theta $, $a=coshr + sinhrcos\theta$, and
$b=coshr - sinhrcos\theta$. At the singularity $r \rightarrow
\varepsilon $, whilst
$\theta $ is free and non-zero,
and hence $ u     \rightarrow
\epsilon sin\theta  $, $a-b \rightarrow  2\epsilon cos\theta $.
Thus
the gauge potential (\ref{sing}) becomes that of a
{\it singular} gauge transformation,
\be
A_{z} \rightarrow \epsilon ^2 \partial _z  \theta
\label{lim}
\ee

\nk which has the interpretation of an $U(1)$
{\it gauge monopole}.
Its topologically non-trivial
behaviour follows from its
angular
dependence, as is
apparent from the
embedding (\ref{emb}). The argument of $z$
winds
non-trivially
around the compact dimension $\theta$, thereby
producing a cut in the two-dimensional manifold
with the form
of
a Dirac string.
Thus
the interpretation of the world-sheet
vortex configuration leading to a Euclidean black hole
as a topologically non-trivial
$U(1)$ gauge field configuration in the $WZ$ model.
\pr
{\it ``Spikes''}: These are
solutions $X_m$ of Green function equations of
the following type \cite{cates,thom}:
\be
   \partial _z  {\overline \partial _z} X_m =-\frac{q_m \pi} {2}
[\delta (z-z_1) -\delta (z-z_2)]
\label{monpair}
\ee

\nk which corresponds to a ``spike''
at $z_1$ and an ``antispike'' at $z_2$.
Once again, the compactness of the closed string world-sheet imposes
zero net ``spikiness'', but in this case $X_m$ is non-compact and
hence aperiodic, so the spike charge $q_m$ is not quantized $a$
$priori$ and the spike is a non-topological soliton. After
stereographic projection of the South Pole on the
sphere onto the origin and of the North Pole onto the point at infinity
in the complex plane,
\be
  X_m = q_m Relnz=q_m ln|z|
\label{mon}
\ee

\nk is the $\sigma$-model coordinate solution of (\ref{monpair}).
\pr
    Although the charges $q_m$ of the spikes are not quantized when
they are in isolation, the $q_m$ $are$ quantized in the presence of
vortices on the world-sheet \cite{thom}. This
can be seen by considering the
interaction of a spike at the point $z$ with a vortex at the point
$z'$ at finite temperature $T$ = $\beta ^{-1}$:
\be
        -\beta S_{eff}(z,z')_{int} =
 2i \beta \pi q_v q_m Imln (z-z')
\label{int}
\ee

\nk The imaginary nature of the effective action (\ref{int})
can be understood
from the observation that the vortex (\ref{vorpair}) and spike
(\ref{monpair}) equations
are related by analytic continuation (Wick rotation) in the
respective charges $q_{v,m}$. Single-valuedness of the partition
function requires that the phase on the right-hand side of
(\ref{int}) also be
single-valued, and therefore imposes the following quantization
condition on the spike charge:
\be
  2\pi \beta q_v q_m =integer
\label{qumon}
\ee

\nk The dual nature of vortices and spikes suggested in (\ref{qumon})
is discussed in ref. \cite{thom} and in the next section.
For the physical application to black holes that is of interest to
us, the minimal vortex charge $q_v=1$ is allowed, so the quantization
condition (\ref{qumon}) becomes
\be
    2\pi \beta q_m \equiv  e = integer
\label{qum}
\ee

\nk at finite temperature $T  =  \beta ^{-1}   \ne 0$.
\pr
    In our application, spikes are to be interpreted as Minkowski
black holes. This is to be expected from the fact that spikes are
non-compact configurations on the world-sheet. To see this
one again has to consider an embedding of the world-sheet
in target space-time which is consistent with conformal invariance.
It turns out that the appropriate one is
\be
|z|^2 = -uv  \qquad : \qquad u=e^{R +t} , v=-e^{R-t}
\label{embmin}
\ee

\nk where $R$ is a spike configuration of unit charge ,
$R=ln|z|$. Taking into account the stereographic projection
of the sphere onto the complex plane,
$R$ expresses a spike-antispike pair located at the origin
(South Pole) and
infinity (North
Pole) respectively.
The target space-metric is that of a Minkowski black hole .
In terms of the embedding in $R,t$ non-compact coordinates
it reads:
\be
    ds^2 =\frac{dzd{\bar z}}{1+z{\bar z}}=
    dr^2 -tanh^2 r dt^2 \qquad : \qquad R=r+ln(1-e^{-2r})
\label{bhmin}
\ee

\nk The coordinate singularity at $r=0$, which corresponds
to the position of the
spike
at $z=0$ on the world-sheet, is
the horizon of the black hole. In this way spikes on the
world-sheet are responsible for generating event horizons
in a Minkowski
space-time. To pass the horizon one needs analytic
continuation in the Kruskal-Szekeres sense \cite{wheel}. This
can be achieved with the
embedding (\ref{embmin}) but using the $u,v$ coordinates,
where the target metric takes the form
\be
ds^2 =-\frac{dudv}{1-uv}
\label{ks}
\ee

\nk These are actually the natural $\sigma$-model
coordinates appearing in the $WZ$ description,
which makes use of
gauging the
non-compact subgroup $O(1,1)$ of $SL(2,R)$ \cite{witt}.
\pr
The
latter
provides
a consistent description of the singularity
using a well-defined topological field theory model.
Reparametrizing the singularity by $w =lnu=-lnv$
one ends up with a topological gauge theory
on the world-sheet \cite{witt}
\be
 S_{CS}= i\int d^2z \sqrt{h}
 \frac{k}{2\pi} w\varepsilon^{ij}(F(A))_{ij}+ ...
\label{topol}
\ee

\nk where $h$ is the (non-trivial) world-sheet metric,
$A_{i}, i=1,2 $ is the
Abelian gauge field of the non-compact
$O(1,1)$ subgroup,
$F_{ij}$ is its field strength, and the
dots represent additional ``matter'' fields
corresponding to space-time coordinates.
Their generic form close to the singularity
is \cite{witt}
\be
     -\frac{k}{4\pi} \int d^2 z\sqrt{h}
h^{ij} D_i a D_j b + (higher-order-terms)
\label{matter}
\ee

\nk where $D_i = \partial _i + iA_i , i=1,2$ is a covariant
derivative, and
$ab + uv=1$.
\pr
In the neighbourhood of the singularity it can be shown using the
equations of motion
that
the gauge field $A$ has the form
\be
\nonumber A_z = -A_{{\bar z}} = -\frac{\partial _z w}{2(1-uv)}
\label{minpot}
\ee

\nk Its analyticity properties depend therefore
on the nature of the $w$ field. We shall now
argue that the relevant configuration is a $w$-{\it monopole}
where the $w$ field is singular, in which case the total
flux of the $A$ field through the compact world-sheet
surface vanishes. To this end, we first note that
the topological term (\ref{topol})
can be thought of
as the dimensional
reduction of a three-dimensional Abelian
Chern-Simons term
\be
i\int _{\Sigma \times S^1}  d^3x
          A(z,{\bar z}, \tau) \wedge F(A(z,
{\bar z},\tau))
\label{chern}
\ee

\nk where
$S^1$ is
the manifold of
a homotopic extension variable $\tau$ which we will use later
as a `pseudo-temperature' to describe the phase structure
of space-time, and
to parametrize adiabatic flow in an ensemble of Minkowski
black hole models. The system (\ref{chern}) becomes
effectively two-dimensional when an adiabatic condition
of periodicity in $\tau$
is imposed on the fields $a$, $b$ and $A$ and the circumference
$2\pi \beta$ of the homotopic $S^1$ is allowed to become
large \footnote{As explained in the next section it is the
low-temperature phase that allows for free Minkowski
black holes which have the non-compact gauge description
(\ref{topol}), so this adiabatic picture is appropriate.
It should be noticed
that
dimensional reduction also occurs in
effective lagrangians of {\it hot} {\it local}
field theories. However this situation is not applicable
to string theory due to
the existence of
winding modes \cite{atick}. Hence our
adiabatic interpretation seems more
appropriate in this context.}.
The
singular
parameter $w$
is related to $A(z,{\bar z}, \tau)_\tau
 \equiv A(\Sigma , \tau)_\tau $
but there is no
formal reason to assume periodicity in $\tau$ for the
homotopically extended $A(\Sigma, \tau)_\tau $.
The $w$ field
which appears
in (\ref{topol}) is viewed as
a $\beta $-independent
ensemble average,
$\frac{1}{\beta}\int_0^\beta d\tau A_\tau $.
The $A_\tau $ field
is allowed to have non-trivial boundary conditions
of the monopole type
$A(\Sigma, \beta )
= A(\Sigma ,0) + \int _0^{\beta} d\tau \partial _\tau \theta
=A(\Sigma ,0) + \Lambda  $, and
the coefficient of the three-dimensional
Chern-Simons term
is a $\beta $-dependent
parameter, ${\tilde k}$, such that  ${\tilde k}\beta$=const=$k$.
This
is allowed
in Abelian gauge theories,
since there is no
{\it a priori} quantisation condition.
Consequently, the Chern-Simons term (\ref{topol})
is not invariant under large gauge transformations but
changes as
\be
 \delta S_{CS} =i\frac{{\tilde k}}{ \pi} \Lambda
 \int _{\Sigma} \epsilon_{ij}\partial _iA_j
\label{change}
\ee

\nk In a path integral formalism, quantum fluctuations
of monopoles are not consistent with fixed boundary
conditions \cite{monopol}, so one has to integrate
over the whole range of the gauge parameter
$\Lambda$. In our
case the gauge field $A_\tau $ might be considered as the
{\it continuum }
version of a corresponding phase variable defined on a discretized
three-dimensional manifold. The circle $S^1$ is viewed as
a discretized real line with spacing $\epsilon$  and periodic
boundary conditions.
Thus
$\Lambda$
takes
values on the interval $[-\frac{\pi}{\epsilon}, \frac{\pi}{\epsilon}]$,
so in the limit $\epsilon \rightarrow 0$ it extends
over the whole range of real numbers. Hence,
one gets the following contribution to the path-integral
\be
 \int _{-\infty}^{+\infty} D\Lambda
 e^{i\frac{{\tilde k}}{2\pi}\Lambda \int _{\Sigma}
\epsilon _{ij}F_{ij}} =\delta [\frac{{\tilde k}}{\pi}\int _{\Sigma} dA ]
\label{flux}
\ee

\nk which implies that
configurations of $A$ with non-zero
total flux through the compact
world-sheet surface are suppressed.
\pr
This result has another elegant intepretation
in terms of self-dual Chern-Simons Higgs models \cite{jackiw},
to which the theory (\ref{topol}),(\ref{matter})
can be mapped. As shown in \cite{emnp}, the Minkowski
black hole corresponds to a self-dual non-topological
soliton in the Bogomolny limit \footnote{This is
consistent with its extreme Reissner Nordstrom character
and justifies the `no-net-force' condition\cite{kal}
used later in our discussion of multi-defect solutions
and space-time foam.}.
In this limit, the magnetic field of the Chern-Simons
gauge potential $A$ vanishes \cite{jackiw,emnp} in the
symmetric vacuum which corresponds to the singularity
\cite{emntop}. Moreover, the Bogomolny limit corresponds
to $N=2$ supersymmetry \cite{oliv,lee} and hence
to the interpretation \cite{witt,eguchi} of target-space
singularities as (twisted) $N=2$ supersymmetric fixed points,
which
leads to an enhancement \cite{emntop}
of the symmetry
of the target-space theory.
\pr
To recapitulate: there are
descriptions of two dual types of
world-sheet
defects, interpreted
as Euclidean and Minkowski black holes,
as
gauge monopoles.
In the Euclidean case
there are monopole-like
configurations of the $A$ field of the $WZ$ model
that describe
coordinate singularities. In the
Minkowski case,
a topological
gauge field theory description of the singularity
is available  in the form of a mixed Chern-Simons
term. In this case there are no non-trivial configurations
for the $A$ field, but still there is a monopole-like
gauge potential for the description of
the singularity,
that of the $w$ field.

\section{The Phase Structure of Space-Time}
\pr
    For the physical applications to the phase structure of target
space-time and to the nature of space-time foam which we make in this
and subsequent sections, we need to study the statistical
distribution of multiple world-sheet defects. This can be derived from
an action principle associated with the following partition function,
where we have made a stereographic projection of the sphere of radius
$r$ onto the complex plane, inducing an effective metric $g(z)$
defined in \cite{thom}:
\be
Z=\int D{\tilde X} exp(-\beta S_{eff}({\tilde X}) )
\label{act}
\ee

\nk where ${\tilde X} \equiv   \beta^{\frac{1}{2}}X$, and
\bea
\nonumber
\beta S_{eff} &=& \int d^2 z [ 2\partial {\tilde X}
{\overline \partial } {\tilde X} +  \frac{1}{4\pi }
[ \gamma _v\epsilon ^{\frac{\alpha}{2}-2}
(2 \sqrt{|g(z)|})^{1-\frac{\alpha}{4}}: cos (\sqrt{2\pi \alpha }
[{\tilde X}(z) + {\tilde X}({\bar z})]):   \\
&+&  (\gamma _v, \alpha,
{\tilde X}(z) + {\tilde X}({\bar z}) )
\rightarrow (
\gamma _m, \alpha ', {\tilde X}(z) - {\tilde X}({\bar z}))]]
\label{eff}
\eea

\nk Here $\gamma_{v,m}$ are the fugacities for vortices and spikes
respectively, and
\be
 \alpha \equiv 2\pi \beta q_v^2  \qquad \alpha ' \equiv
\frac{e^2}{2\pi \beta}
\label{anom}
\ee

\nk are related to the conformal
dimensions $\Delta_{v,m}$ of the vortex and
spike creation operators respectively, namely
\bea
\nonumber
\alpha =4 \Delta _v \qquad   \alpha ' =4 \Delta _m \\
      \Delta _m =\frac{(eq_v)^2 }{16 \Delta _v}
\label{conf}
\eea

\nk It is
easily seen \cite{thom} that the effective action (\ref{eff})
is {\it invariant} under the simple
duality transformation
$ q_v   \leftarrow \rightarrow e $, $\pi \beta    \leftarrow \rightarrow
           \frac{1}{4\pi \beta}$,
which is thus an {\it exact} symmetry of the system.
Indeed, systems of the form (\ref{eff}) are known
to possess \cite{rab}
extended (complex) duality symmetries $SL(2,{\bf Z})$
which could explain
the quantum Hall effect hierarchy \cite{lutk}.
As discussed in \cite{emnp}, the Minkowski
black hole case is field-theoretically
equivalent to a Hall conductor, motivating
further studies of the
implications of such world-sheet
symmetries
for the physics of
black holes. We also note
the connection of the cosine terms
in (\ref{eff}) with the gauge field theory description
of the defects discussed in the previous section. For
instance, in the case of Minkowski black holes, the cosine
terms in (\ref{eff}) correspond after fermionization
\cite{colemanth} to
interactions of a massive Thirring
model. A similar result holds for the fermionized
Chern-Simons lagrangian coupled to scalar Higgs
fields around
the asymmetric vacuum \cite{ruiz}, which in our
case corresponds to
the Minkowski black hole
horizon (position of the spike
in the embedding (\ref{embmin})). In this case,
the Chern-Simons term can be represented by an
infinity of higher order fermion interactions,
of which the four-fermion one is a relevant operator
in the renormalisation group sense \cite{ruiz,emnp}.
\pr
Since the relevant dynamics of the partition function
(\ref{act}) is that
on the two-dimensional world-sheet, only generalized
Berezinsky-Kosterlitz-Thouless phase transitions \cite{KT}
are possible. These may be
induced by either vortices or spikes (or both) when the corresponding
anomalous dimension(s) become equal to unity. When
$\Delta_{v,m}$ $>$ 1 the corresponding charges are
confined
and
screened,
when $\Delta_{v,m}$ = 1 the corresponding operator is marginal and a
phase transition occurs, and when $\Delta_{v,m}$ $<$ 1 it is relevant and
the corresponding charges dissociate to form a plasma. Such phases
have been discussed previously in other applications, but the
physical interpretation here as the phase structure of space-time is new.
\pr
    We see from the expression (\ref{conf}) for the conformal
dimension $\Delta_m$ that free
spikes
occur when the
temperature
\be
         T <       T^c_m  =  (\beta^c_m)^{-1} =
         \frac{8\pi}{e^2}
\label{twelve}
\ee

\nk and we interpret
this phase as containing a plasma of free Minkowski
black holes. The corresponding expression for the conformal dimension
$\Delta_v$ tells us that free vortices occur when the temperature
\be
         T >         T^c_v  =  (\beta^c_v)^{-1} =
\frac{\pi q_v^2}{2}
\label{thriteen}
\ee

\nk and we interpret this
phase as containing a plasma of free Euclidean
black holes : $T^c_v$ is conventionally
known as the Hagedorn temperature \cite{sath}.
On the other hand, we see that when $T$ $>$ $T^c_m$ the
Minkowski black holes (spikes) are
confined, whereas the Euclidean black
holes (vortices) are confined when $T$ $<$ $T^c_v$. There are three
possible phase diagrams depending on whether $eq_v $ $>$, $=$
or $<$ $4$.
\pr
In each case there
is a low-temperature phase, in which we live, in which Minkowski
black holes form a free plasma and Euclidean black holes (vortices)
are confined, and a high-temperature phase, possibly realized in the very
early Universe, in which Euclidean black holes form a free plasma and
Minkowski black holes (spikes) are confined. In this sense
space-time could have seemed Euclidean in the early Universe. We know
from standard results on the renormalization group behaviour of
vortex/spike systems that whereas the vortex fugacity remains small
in the low-temperature phase, it is driven to large values in the
high-temperature phase. This means that the high-temperature phase is
a {\it dense} vortex plasma. Since the core of a Euclidean black hole is
characterized by a compact $U(1)$ topological gauge field theory on
the world-sheet, which has a higher super-$W$ symmetry with a bosonic
$W_{1+\infty} \otimes  W_{1+\infty}$
subsymmetry \cite{emntop}, we expect that this symmetry
would have been present in this early
dense high-temperature phase.
This expectation is supported by the observation \cite{atick}
that the number of effective degrees of freedom at high
temperatures is characteristic of a two-dimensional
field theory.
\pr
    In the case $eq_v$ $>$ 4, in addition to this
high-temperature phase and the familiar low-temperature phase with
free Minkowski black holes and confined vortices, there is an
intermediate range of temperatures $\beta^c_v$ $<$ $\beta$
$<$ $\beta^c_m$
in which neither type of black hole, i.e., neither free vortices nor
free spikes, exist, but both vortices and spikes are
confined. This
intermediate phase disappears when $eq_v$ = 4.
Finally,
when $eq_v$ $<$ 4
a new intermediate
phase appears when $\beta^c_m$ $<$ $\beta$ $<$
$\beta^c_v$, in which black
holes of both Minkowski and Euclidean types coexist freely, i.e.,
neither vortices nor spikes are
confined. In each of these cases, the
appearance of a vortex condensate breaks the bosonic
$W_{1+\infty}\otimes W_{1+\infty}$ subsymmetry of the initial
super-$W_{1+\infty}$ down to a residual single $W_{1+\infty}$,
responsible for the maintenance of quantum coherence \cite{emn1},
and
discrete massless modes appear \cite{emntop}, like
pions in QCD below the
quark-hadron phase transition.
\pr
Which of $eq_v$ $>$, $=$, $<$ $4$ occurs
in Nature? By analogy with the $XY$
model, one would expect the dominant vortex configuration to be that
with $q_v$ = 1. This expectation is confirmed by the following simple
free energy argument. In the stringy space-time interpretation of the
vortex, the Euclidean time variable is identified with a conformal
field ${\tilde X} $ compactified on a circle, that describes the
``matter'' part
of a Liouville-matter theory, with kinetic term normalized to
\be
\frac{1}{2\pi}
\int d^2 z \partial {\tilde X} {\overline \partial }{\tilde X}
 \label{confo}
\ee

\nk The
energy of an isolated vortex (Euclidean black hole) of charge
$q_v$ is then $\pi q_v^2 ln(\frac{R}{a})$, where $R$
is the size of the system
and $a$ is
an ultraviolet cutoff, and its entropy is $ 2ln(\frac{R}{a})$,
where the prefactor is determined by the two-dimensionality of the
world-sheet. Hence the free energy of an isolated vortex is
\be
   f_s = (\pi q_v^2 -2)ln(R/a)
\label{energ}
\ee

\nk indicating that a
vortex plasma is energetically preferred
when $q_v = 1$. A similar result
holds for the spike configurations of the Liouville field $\phi$ of
quantum gravity. The dominant terms in the effective action for this
field have the form
\be
      S_L = \frac{25-D}{96\pi} \int d^2z[(\nabla \phi)^2 +
        \phi R^{(2)} + ...]
\label{liouv}
\ee

\nk where $D$
is the central charge of the matter theory, i.e., the target
space dimensionality, and the dots represent deviations from flat
target space backgrounds whose detailed form does not affect our
arguments. Note the unconventional $D$-dependent normalization of the
kinetic term in (\ref{liouv}). In
the string case with $D$ $>$ 25, one should
continue analytically $\phi$ to $i\phi$ and view the Wick-rotated $\phi$
as proportional to target time. Recalling that the Euler
characteristic of the sphere is
$\int R^{(2)} =8\pi$, we observe that the action
(\ref{liouv}) is periodic for
$\phi \rightarrow  \frac{12n}{25-D} \phi$,
with $n$ an integer. Thus a convenient
normalization to discuss spike solutions in Liouville theory,
corresponding to a Wick rotation of vortices, is
\be
{\tilde \phi} \equiv \frac{25-D}{12}\phi
\label{redf}
\ee

\nk
One can then introduce a pseudo-temperature
\be
     T_{ps}  =  (\beta_{ps})^{-1} \qquad : \qquad \beta_{ps}  =
\frac{3}{\pi ( D  - 25 )}
\label{eighteen}
\ee

\nk in terms of which one has the following quantization condition
analogous to (\ref{qum}) :
\be
        2\pi\beta _{ps} q_m =e=integer
\label{strinqu}
\ee

\nk Taking
account of the unconventional normalization of the spike
field, we see that the self-energy of such a spike is
\be
          U=\frac{3}{(25-D)} q_m^2 ln(R/a)
\label{uen}
\ee

\nk where $R$ is
the size and $a$ the ultraviolet cutoff of the system as
in the vortex case, and the entropy of an isolated spike is also
$ 2ln( R / a )$. Using the quantization condition
(\ref{strinqu}), we see that the
free energy is again minimized for the physical case
$D = 1 $
when $e$ = 1. Taking into account the normalization
(\ref{redf}), we see
that this corresponds to the charge-2 spikes of Cates \cite{cates}.
We conclude that the lowest-lying vortex and spike configurations are
\be
              q_v = e = 1
\label{phys}
\ee

\nk implying that
the case $eq_v$ $<$ $4$ describes the black hole problem.
There is an intermediate phase where free Euclidean (vortex)
and Minkowski (spike) black holes $both$ occur.
\pr
    It should be stressed that the above analysis applies to
Minkowski black holes of arbitrary mass. This can be seen formally by
considering the case when the background metric in (\ref{liouv})
is flat but
with an overall scale factor $e^{-2\phi _0}$, corresponding to a constant
shift in the dilaton.
This factor expresses the (dimensionless)
ratio of the
black hole mass
renormalized by the pseudo-temperature
to an initial scale set by the choice of the
constant dilaton shift.
The above arguments are easily modified by
introducing factors of $e^{2\phi _0}$ into the definition  of
$\beta_{ps}^{-1}$ (\ref{eighteen})
and $q_m$ (\ref{mon}).
\pr
We would also like to stress that the above arguments for
the three
phases of the $eq_v $ $<$ $4$ case, a
high-temperature phase with a dense plasma
of Euclidean and confined Minkowski black holes, an
intermediate-temperature phase with both types free, and a
low-temperature phase with a plasma of Minkowski black holes and
confined Euclidean black holes that breaks a higher symmetry,
are likely to apply to the physical situation of three target space
dimensions. The arguments that the dominant vortex and spike
configurations would be those with $q_v$ = $e$ = 1 do not depend on
the dimensionality of target space, but only
on the fixed
total central charge
in the case of the
spike.
The self-energy of
any individual vortex or spike
is determined by properties of the
world-sheet, notably its two-dimensionality. The numbers of embeddings
of a vortex or spike configuration in the physical space would have
constant (i.e., $R$- and $a$-independent) extra numerical factors, and
so would not affect the dominant logarithmic behaviour in the entropy,
which is also determined by the dimensionality of the world-sheet.
Therefore the above naive free-energy arguments on the circumstances
when vortex or spike condensation occurs go through for any dimension
of target space.  In such a case, the relevant expression
for the free energy of an isolated spike would read
\be
       f_D= 2[\frac{e^2}{8\pi\beta_{ps} (D)}-1 ]ln(R/a)
\label{dimens}
\ee

\nk and obviously corresponds (c.f.
(\ref{eighteen}) after
the replacement $D-25$ $\rightarrow$ $ 25-D$)
to a pseudo-temperature smaller
than the critical one ($D=1$), and, therefore, to
the region of the phase-diagram where there
is a proliferation
of spikes as we discussed above.
It should be stressed, though,
that these arguments, although suggestive,
do not prove
the
existence of
exact conformal field theory models
for the description
of higher-dimensional strings  ($1$ $<$ $D$ $<$ 25)
in singular space-time
backgrounds.

\section{The Quantum Fate of a Minkowski Black Hole}
\pr
    In the light of the above results, we are now able to discuss
the quantum evolution of a Minkowski black hole
induced by
higher-genus corrections to the effective action, and its quantum
fate, which we argue is to become lost among the quantum
fluctuations intrinsic to space-time foam. We first recall that the
mass of a Minkowski black hole is given by \cite{witt,wheel}
\be
                    M_{bh} =\sqrt{\frac{2}{k-2}}e^{\phi _0}
\label{twentytwo}
\ee

\nk where $k$ is the level of the $SL(2,R)/O(1,1)$ coset model, which is
related to the central charge: $c  = \frac{3k}{k-2}$  so
that $c$ = 26
corresponds to $k$ = 9/4, $\phi _0$ is a constant of integration that
reflects the asymptotic value of the dilaton field
\be
                   \Phi  = 2ln cosh(r) + \phi _0
\label{twentythree}
\ee

\nk and the metric is
\be
ds^2  = e^{-2\phi _0}[dr^2  -  tanh^2rd\theta^2]
\label{metr}
\ee

\nk We have argued
previously \cite{emn3} that a Minkowski black hole is
classically stable for any value of $\phi _0$. Particle
emission from black
hole is an intrinsically quantum phenomenon that appears in string
theory only when higher-genus effects are taken into account, and we
have displayed explicitly the imaginary part of the one-loop effective
action of the $SL(2,R)/O(1,1)$ coset model
that corresponds to non-thermal
Minkowski black hole decay in any space-time dimension \cite{emn3}.
\pr
    Since perturbative corrections cannot change the total central
charge $c$, higher-genus decay
effects cannot change the level $k$, but can
renormalize the effective value of the integration constant $\phi _0$.
Physically, we would expect in the light of the mass formula
(\ref{twentytwo})
that they would tend to decrease $\phi _0 $. In the limit of large $r$ at
fixed $\phi _0$, or at fixed $r$ for $\phi _0 \rightarrow -\infty $, the
dilaton field $\phi$ (\ref{twentythree})
becomes linear in $r$, and the metric (\ref{metr})
resembles that of flat
space. Thus the Minkowski black hole looks asymptotically like the
flat $c$ = 1 string model.
\pr
The dynamics can be described using
the three-dimensional homotopic extension of the
effective gauge theory introduced in section 2. The r\^ ole
of the radius of the extra dimension, the homotopic scale
$\beta $, provides us with a scale whose flow determines
the black hole decay, in analogy with
the finite-size scaling \cite{car} of two-dimensional
systems formulated on a strip of width $L$. Away from
renormalisation group criticality \cite{mavmir},
the free energy of the finite-size system is related
to the ``bulk'' free energy (corresponding to
$L \rightarrow \infty$) through the relation
\be
F_L = F_{bulk} + \frac{V}{L^2}[C(g) + O(\epsilon ln L)]
\label{finite}
\ee

\nk where $V$ is the strip volume, $\epsilon$ is a covariant
ultraviolet
cut-off, \{ g \} is a set of coupling constants,
and $C(g)$ is related to the Zamolodchikov $c$-function
\cite{zam}, which
at the fixed points becomes the central charge of the theory.
Relation (\ref{finite}) defines an `effective' central
charge which determines the critical properties of the
finite-size system \cite{mavmir}. In our case. the
r\^ ole of the scale $L$ is played by the homotopic scale
$\beta$
leading to a
definition of an effective central charge
in the Liouville sector which
varies
with the homotopic scale $\beta$.
This Liouville-sector central charge deficit is
compensated by an equal and opposite variation
in the matter sector
of the theory at {\it higher genera}, so as to maintain the
conformal invariance ($c_{total}=26$) of the complete
system.
It should be stressed that although this balancing of
the central charges in the Liouville and matter sectors
is familiar from the theory of non-critical strings \cite{aben},
in our case there is an important difference. In the
generic non-critical string theory with $D$ $>$ $25$,
this balance occurs already at string tree level due to
ordinary tachyonic instabilities related to
relevant operators to drive the flow. In our case, it is
a higher genus effect that drive the flow.
\pr
This is
easily seen by the observation \cite{pech} that
non-critical string theory formulated on a world-sheet torus
can be consistently regularised with respect to {\it both}
modular and ultraviolet infinities at the cost of
introducing ``spiky'' configurations in the dilaton
field. The generic result of \cite{pech} is that a covariant
heat-kernel
regularization procedure
with a cut-off $\epsilon$
yields for the free string
partition function in $d$ space-time dimensions
\be
  Z_{torus} \propto
  (\frac{V}{4\pi \epsilon})^{-\frac{1+d}{2}}
exp[-(\frac{1}{2}d -1 )\frac{V}{4\pi \epsilon}\Psi (\alpha _c)]
\label{renorm}
\ee

\nk where $\Psi (\alpha _c)$ is a function evaluated at
a certain saddle point $\alpha _c $,
whose detailed form is not important
for our subsequent arguments.
In two dimensions the tachyonic infinities are absent but there
are logarithmic infinities (`spikes' $e^{ (1+\frac{d}{2})ln\epsilon}$)
in the torus partition function which cannot be absorbed by
the usual renormalization of the torus
background coupling constants
of the $\sigma$-model. They are
absorbed by renormalization of the
$\sigma$-model couplings on the sphere \cite{fischl}.
They are
divergent
contributions to the dilaton field that couples to the Euler
characteristic of the sphere. This produces a renormalization
of the conformal `anomaly' of the non-critical string in both
the matter and the Liouville sectors \cite{pech}. This enables
higher-genus instabilities of the black hole space-time
to be
incorporated consistently
in a flow
where the total cental charge remains zero,
and
we know this happens thanks to the
exact conformal invariance of the matrix model.
This discussion clearly applies
only
to the non-compact Minkowski case, which
is the only one that can allow variation of the black hole mass.
This picture is perfectly consistent with the one advocated
in \cite{emn3} from a two-dimensional string effective action
($S$-matrix amplitude) point of view.
\pr
This balance of central charges between the Liouville and matter
sectors during the evaporation/decay
of the stringy black hole constitutes precisely
what in general relativity would be called the
{\it back reaction}
of quantum effects on the gravitational dynamics.
The irreversibility of black hole decay
can be understood \cite{emnp} by
analogy with the isomorphic theory
of the Quantum Hall Effect
as a manifestation
of the violation of parity and time-reversal induced
by the three-dimensional Chern-Simons term (\ref{chern}).
When
applied to the initial
singularity, with the r\^ ole of target
time
played
by the Liouville mode, such a picture
would provide an explanation of the arrow of target time as well.

\section{Conclusions and Prospects}
\pr
We have shown in this paper how the two dual species
of defects on the world-sheet, vortices and spikes,
are related to black holes of Euclidean and Minkowski
signature respectively, and in both cases related to
gauge defects in the $WZ$ coset model descriptions.
Taking over known results on Berezinsky-Kosterlitz-Thouless
phase transitions, we have argued that space-time
has three phases : a high-temperature phase
with a Euclidean black hole (vortex) plasma
and confined Minkowski black holes (spikes),
an intermediate temperature phase with free black holes
of both signatures, and a low-temperature phase with a
plasma of Minkowski black holes (spikes)
and confined Euclidean black holes (vortices).
This last phase constitutes stringy space-time foam.
Thermal Euclidean black holes have quantized
winding number and cannot decay, whilst Minkowski
black holes have a continuous mass spectrum
and decay via coherent higher-genus quantum corrections
which cause a back-reaction that
reduces their masses until they become indistinguishable
from the quantum-mechanical fluctuations intrinsic to space-time
foam. Thus, in our view the quantum fate of a black hole
is not to leave behind a stable remnant, nor to disappear
completely, but ``none of the above '', namely to become
a nondescript quantum fluctuation in the vacuum.
\pr
The phase structure discussed above was derived in the case
of a two-dimensional target space-time, but we have given
arguments why it could also apply to the relevant
four-dimensional case. Certainly,
our understanding of black hole decay and back-reaction
as coherent quantum phenomena applies to (at least)
spherically-symmetric black holes in any number of dimensions.
Nevertheless, it is desirable to acquire a deeper understanding
of space-time foam in four dimensions, as well as the nature
of the phase diagram in this case. A deeper understanding
of the high-temperature phase with space-time symmetry
restoration is also desirable, together with its
embedding in a realistic cosmological context.
\pr
We are encouraged by the successes of string theory
in providing solutions for many of the key problems
in quantum gravity to believe that these
outstanding questions now are also vulnerable to answers
from string theory.
\pr
\noindent {\Large{\bf Acknowledgements}} \\
\par
The work of D.V.N. is partially supported by DOE grant
DE-FG05-91-ER-40633 and by a grant from Conoco Inc.
\pr


\begin{thebibliography}{99}
\bibitem{mand} S. Mandelstam, Phys. Lett. B277 (1992), 82.
\bibitem{witt} E. Witten, Phys. Rev. D44 (1991), 314.
\bibitem{emn4} J. Ellis, N.E. Mavromatos and D.V. Nanopoulos,
Phys. Lett. B278 (1992), 246.
\bibitem{emn1} J. Ellis, N.E. Mavromatos and D.V. Nanopoulos,
Phys. Lett. B267 (1991), 465;
Phys. Lett. B272 (1991), 261.
\bibitem{ind} S. Das, A. Dhar, G. Mandal, S. Wadia,
ETH-TH-91/30; IASSNS-HEP-91-52; TIFR-TH-91-44.
\bibitem{whoW} I. Bakas, Phys. Lett. B228 (1989), 57;
\par C.N. Pope, X. Shen and L.J. Romans, Nucl. Phys. B339
(1990), 191.
\bibitem{emntop} J. Ellis, N. E. Mavromatos and
D.V. Nanopoulos, CERN-Texas A \& M Univ.
preprint
CERN-TH.6514/92; ACT-11/92; CTP-TAMU-43/92.
\bibitem{KT} V.L. Berezinsky, Zh. Eksp. Teor. Fiz.
59 (1970), 907; 61 (1971), 1144;
\par J.M. Kosterlitz and D.J. Thouless, J. Phys. C6,
(1973) 1181.
\bibitem{eguchi} T. Eguchi, Mod. Phys. Lett. A7 (1992), 85.
\bibitem{chiral} B. Campbell,
J. Ellis, S. Kalara, D.V. Nanopoulos and K. Olive,
Phys. Lett. B255 (1991), 420.
\bibitem{emn3} J. Ellis, N.E. Mavromatos and D.V. Nanopoulos,
Phys. Lett. B276 (1992), 56.
\bibitem{abr} A.A. Abrikosov, Jr., and Ya.I. Kogan,
Int. J. Mod. Phys. A6 (1991), 1501.
\bibitem{thom} B. Ovrut and S. Thomas, Mod. Phys. Lett. A5 (1990),
2351; Phys. Rev. D43 (1990), 1314; Phys. Lett. B257 (1991), 292.
\bibitem{chinese} H.B. Gao and Y.X. Chen, Zhejiang University
preprint ZIMP-91-23 (1991).
\bibitem{cates} M.E. Cates, Europhys. Lett. 8 (1988), 719.
\bibitem{wheel} C.W. Misner, K.S. Thorne and
J.A. Wheeler, {\it Gravitation} (W.H. Freeman and Co., 1973).
\bibitem{atick} J.J. Atick and E. Witten, Nucl. Phys. B310
(1988), 291.
\bibitem{monopol} I. Affleck, J. Harvey, L. Palla and G. Semenoff,
Nucl. Phys. B328 (1989), 575, and references therein.
\bibitem{jackiw} R. Jackiw and E. Weinberg, Phys. Rev.
Lett. 64 (1990), 2234;
\par R. Jackiw, K. Lee and E. Weinberg, Phys. Rev. D42 (1990),
3488.
\bibitem{emnp} J. Ellis, N.E. Mavromatos and D.V. Nanopoulos,
CERN-Texas A \& M Univ. preprint, CERN-TH.6536/92; ACT-13/92;
CTP-TAMU 48/92 (1992).
\bibitem{kal} S. Kalara and D.V. Nanopoulos, Phys. Lett.
B267 (1991), 343, and references therein.
\bibitem{oliv} E. Witten and D. Olive, Phys. Lett. B78 (1978), 97.
\bibitem{lee} C. Lee, K. Lee and E. Weinberg, Phys. Lett.
B243 (1990), 105.
\bibitem{rab} J. L. Cardy and E. Rabinovici, Nucl. Phys.
B205 (1982), 1;
\par J.L. Cardy, Nucl. Phys. B205 (1982), 17.
\bibitem{lutk} C.A. Lutken and G. G. Ross, Phys. Rev. B45 (1992),
to appear.
\bibitem{colemanth} S. Coleman, Phys. Rev. D11 (1975), 2088.
\bibitem{ruiz} N.E. Mavromatos and M. Ruiz-Altaba, Phys. Lett.
A142 (1989), 413.
\bibitem{sath} B. Sathiapalan, Phys. Rev. D35 (1987), 3227;
\par Ya. I. Kogan, Pis'ma Zh. Eksp. Teor. Fiz. 45
(1987), 556.
\bibitem{car} H.W. Blote, J.L. Cardy and M.P. Nightingale,
Phys. Rev. Lett. 56 (1986), 742;
\par I. Affleck, Phys. Rev. Lett. 56 (1986), 746.
\bibitem{mavmir} N.E. Mavromatos and J.L. Miramontes, Phys. Lett
B226 (1989), 291.
\bibitem{zam} A.B. Zamolodchikov, JETP Lett. 43 (1986), 731.
\bibitem{aben} I. Antoniadis, C. Bachas, J. Ellis and
D.V. Nanopoulos, Phys. Lett. B211 (1988), 393; Nucl. Phys. B328 (1989),
117; Phys. Lett. B257 (1991), 278.
\bibitem{pech} E. Cohen, H. Kluberg-Stern and R. Peschanski,
Nucl. Phys. B328 (1989), 499.
\bibitem{fischl} W. Fischler and L. Susskind, Phys. Lett.
B171 (1986), 383; {\it ibid} B173 (1986), 262.
\end{thebibliography}
\end{document}